\documentstyle[aps]{revtex}

\def\be{\begin{equation}}
\def\ee{\end{equation}}

\def\1{r}
\def\2{\theta}
\def\3{\phi}
\def\4{t}

\begin{document}

\draft

\title{On the strength of the Kerr singularity and cosmic censorship}

\author{%
Wies{\l}aw Rudnicki\thanks{E-mail: rudnicki@atena.univ.rzeszow.pl}
and Pawe{\l} Zi\c{e}ba\thanks{E-mail: pzieba@atena.univ.rzeszow.pl}}

\address{Institute of Physics, University of Rzesz\'ow,
ul. Rejtana 16 A, PL 35-310 Rzesz\'ow, Poland}

\maketitle

\begin{abstract}
It has been suggested by Israel that the Kerr singularity cannot be
strong in the sense of Tipler, for it tends to cause repulsive effects.
We show here that, contrary to that suggestion, nearly all null
geodesics reaching this singularity do in fact terminate in Tipler's
strong curvature singularity. Implications of this result are discussed
in the context of an earlier cosmic censorship theorem which constraints
the occurrence of Kerr-like naked singularities in generic collapse
situations.
\end{abstract}

\section{Introduction}
The cosmic censorship hypothesis of Penrose \cite{penrose69} states
that a physically realistic gravitational collapse can never result in
a ``naked'' singularity---that is, all singularities arising in such
situations should always be enclosed within an event horizon and hence
invisible to distant observers. This hypothesis plays a fundamental
role in the theory of black holes. Unfortunately, in spite of much
sustained effort, no complete proof (or convincing counterexample) to
cosmic censorship has been found as yet. Many partial results, however,
have been established which considerably restrict the class of possible
naked singularities. A~cosmic censorship theorem of this type has
previously been proved by one of the present authors \cite{rudnicki96}.
The aim of this paper is to answer a certain essential question closely
related to that result.

There exist many exact solutions of the Einstein equations in which
naked singularities {\em do} occur. Among these solutions the most
important one, due to its direct astrophysical applications, is
undoubtedly that of Kerr \cite{kerr}. This solution depends
on two parameters $m$ and $a$ (see below), and represents the exterior
gravitational field of a rotating body with mass $m$ and angular
momentum $ma$, as measured from infinity in geometrized units with
$c=G=1$. As is well known \cite{boyer}, the maximal analytic
extension of the Kerr solution with $m>0$ and $a\neq 0$ contains
the ring curvature singularity that may be interpreted as the outcome
of collapse of a rotating object. When $|a|\leq m$, this singularity is
always hidden behind an event horizon. But if $|a|>m$, there is {\em
no} event horizon and the singularity is visible for all observers;
moreover, there are closed timelike curves through every point of the
spacetime \cite{carter}. Clearly, the Kerr solution is highly idealized
and the singularity with $|a|>m$ cannot be a counterexample to
Penrose's hypothesis. However, it is not unlikely that the pathologies
occurring in the case $|a|>m$ could also arise in more general
scenarios of the collapse of a rapidly rotating star.

The censorship theorem of Ref. [2] shows that, under certain reasonable
assumptions, a physically realistic collapse developing from a regular
initial state cannot lead to the formation of a final state resembling
the Kerr solution with $|a|>m$ ---i.e. of a naked singularity
accompanied by closed timelike curves. An important role in this result
plays a certain {\em inextendibility condition}, which is assumed
to hold for all (achronal) null geodesics terminating at the Kerr-like
naked singularity under consideration. This condition characterizes the
curvature strength of the singularity; roughly speaking, it holds for a
given null geodesic $\lambda$ if the curvature near the singularity is
strong enough so that at least one irrotational congruence of Jacobi
fields along $\lambda$ is forced to refocus as $\lambda$ approaches the
singularity (for more details on this condition, see Refs.
\cite{rudnicki95,rudzie}).

It is well known \cite{rudzie} that the inextendibility condition will
always hold for achronal null geodesics terminating at the so-called
{\em strong curvature singularities} defined by Tipler \cite{tipler}
(see below). These singularities have the property that all objects
approaching them are crushed to zero volume. One often assumes that all
singularities arising in physically realistic collapse should be of the
strong curvature type (see, e.g., Refs. [9--11]). However, Israel
\cite{israel} has suggested that the Kerr singularity fails to be
strong in the sense of Tipler's definition, for it tends to cause
repulsive effects. Since the inextendibility condition is similar in
spirit to Tipler's definition (in both cases some Jacobi fields are
refocused), the question one immediately asks is whether this condition
can still be expected to hold for null geodesics terminating at the
Kerr singularity---i.e. whether the censorship theorem of Ref.
\cite{rudnicki96} can be applied to singularities of this type. In
this paper we will obtain a result that provides some positive answer
{}to this question. Namely, we will show that, in general, all null
geodesics reaching the Kerr singularity with
$m>0$ and $a\neq 0$ {\em do} in fact terminate, contrary to the
suggestion of Israel, at Tipler's strong curvature singularity.

\section{Preliminaries}
{}To begin with, we need to recall some of the basic facts on the Kerr
solution. In Boyer and Lindquist coordinates $(r,\theta,\phi,t)$ the
Kerr metric is given by (cf. Ref. \cite{hawking}, p. 161):
\be
ds^2 =
\rho^{2}\left(\frac{dr^{2}}{\Delta}+d\theta^{2}\right)
+ (r^{2}+a^{2})\sin^{2}\theta d\phi^{2} - dt^{2} +
\frac{2mr}{\rho^{2}}(a\sin^{2}\theta d\phi-dt)^{2},
\ee
where $\rho^2 \equiv r^{2} + a^{2}\cos^{2}\theta$ and $\Delta \equiv
r^{2}-2mr+a^{2}$. As mentioned earlier, the constant $m$ represents the
mass of the metric source while $a$ is its angular momentum per unit
mass. The Kerr spacetime is stationary and axisymmetric, with Killing
vector fields $\xi^{a}\equiv (\partial/\partial t)^{a}$ and
$\omega^{a}\equiv (\partial/\partial \phi)^{a}$. Moreover this
spacetime is Ricci flat and is of Petrov type D. The ring singularity
is located in the equatorial plane $\theta=\pi/2$ at points where
$r=0$. As shown by Carter \cite{carter}, the only null geodesics which
can reach this singularity are those lying strictly in the equatorial
plane on the positive $r$ side. The equations of motion for these
geodesics are (cf. Ref. \cite{chandra}, p. 328):
\be
u^{r}\equiv
\frac{dr}{ds}=\pm\left[E^{2}+\frac{2m}{r^{3}}\left(L-aE\right)^{2}-
\frac{1}{r^{2}}\left(L^{2}-a^{2}E^{2}\right)\right]^{1/2},
\ee
\be
u^{\phi}\equiv \frac{d\phi}{ds}=\frac{1}{\Delta}\left[\frac{2ma}{r}E+
\left(1-\frac{2m}{r}\right)L\right],
\ee
\be
u^{t}\equiv \frac{dt}{ds}=\frac{1}{\Delta}\left[\left(r^{2} +
a^{2}+\frac{2a^{2}m}{r}\right)E - \frac{2ma}{r}L\right],
\ee
where $u^{r}$, $u^{\phi}$ and $u^{t}$ are the coordinate basis
components of the tangent vector, ${\bf u}$, to a given null geodesic
$\lambda(s)$ parametrized by an affine parameter $s$. (Note that the
corresponding component $u^{\theta}\equiv d\theta/ds$ of ${\bf u}$
identically vanishes as $\lambda(s)$ lies in the equatorial plane.) The
quantities $E$ and $L$ are the constants of the motion associated with
the Killing vectors $\xi^{a}$ and $\omega^{a}$; they are defined as
follows: $E\equiv -\xi_{a}u^{a}$ and $L\equiv \omega_{a}u^{a}$.
Physically, $E$ and $L$ can be interpreted, respectively, as the energy
at infinity and the angular momentum about the symmetry axis,
$\theta=0$, of a photon moving along $\lambda(s)$. It is also worth
noting here that if $L=aE$, then $\lambda(s)$ must belong to one of the
two principal null congruences associated with the algebraic type D of
the solution (see Ref. \cite{chandra}, p. 329). The $\pm$ signs in Eq.
(2) correspond to outgoing and ingoing geodesics, respectively.

Let us also recall that, according to Tipler's definition
\cite{tipler}, an affinely parametrized null geodesic $\lambda(s)$ is
said to terminate in a {\em strong curvature singularity} at affine
parameter value $s_{0}$ if the following holds (cf. Ref.
\cite{tipler1}, p. 160): Let $\mu(s)$ be a 2-form on the normal
space to the tangent vector to $\lambda(s)$ determined by two linearly
independent vorticity-free Jacobi fields ${\bf Z}_{1}(s)$ and ${\bf
Z}_{2}(s)$ along $\lambda(s)$, i.e. $\mu(s)\equiv {\bf Z}_{1}\wedge
{\bf Z}_{2}$. For {\em all} $\mu(s)$ that vanish for at most finitely
many $s$ in some neighbourhood $(s_{0},s_{1}]$ of $s_{0}$, we have
$\lim_{s\rightarrow s_{0}}||\mu(s)||=0$. Very useful criteria for
determining whether a given null geodesic terminates in a strong
curvature singularity have been found by Clarke and Kr\'olak
\cite{clarke}. One of these criteria, which will be used in proving our
result, can be formulated as the following

{\em Proposition 2.1.} Let $\lambda(s)$ $(0<s\leq s_1)$ be an affinely
parametrized null geodesic and let $\{{\bf E}_{i}\}$ $(i=1,2,3,4)$ be
a pseudo-orthonormal tetrad parallely propagated along $\lambda(s)$,
with ${\bf E}_{1}{\bf \cdot E}_{1}={\bf E}_{2}\cdot {\bf E}_{2}=-{\bf
E}_{3}\cdot {\bf E}_{4}=-{\bf E}_{4}\cdot {\bf E}_{3}=1$, all other
scalar products vanishing and ${\bf E}_{4}={\bf u}$, where ${\bf u}$
is the tangent vector to $\lambda(s)$. Let $C^{m}\null_{4n4}$
$(m,n\in \{1,2\})$ be some component of the Weyl tensor with respect
{}to the tetrad $\{{\bf E}_{i}\}$. If there exists some affine parameter
value $s_0\in (0,s_1]$ such that $C^{m}\null_{4n4}\geq Ks^{-2}$ on
$(0,s_0]$, where $K$ is some positive constant, then $\lambda(s)$
terminates in Tipler's strong curvature singularity as $s\rightarrow
0$.

{\em Proof}. The proof follows immediately from Proposition 7 of Ref.
\cite{clarke}.

\section{The main result}
We are now in a position to state and prove our main result.
\vspace{4mm}

{\em Theorem 3.1.} Let $\lambda(s)$ $(s>0)$ be an affinely parametrized
null geodesic in the Kerr spacetime with $m>0$ and $a\neq 0$. Suppose
that $\lambda(s)$ approaches the ring singularity as $s\rightarrow 0$.
If $\lambda(s)$ does not belong to either of the two principal null
congruences, then $\lambda(s)$ terminates in Tipler's strong
curvature singularity as $s\rightarrow 0$.

\vspace{4mm}

In very brief outline, the proof of our result runs as follows. We
first construct a certain pseudo-orthonormal tetrad $\{{\bf E}_i\}$
parallely propagated along the geodesic $\lambda(s)$. We next find one
of the components of the Weyl tensor with respect to the tetrad $\{{\bf
E}_i\}$. It turns out that if $\lambda(s)$ is not a member of the
principal null congruences, then this component must grow along
$\lambda(s)$ as fast as $\sim r^{-5}$, where $r$ is the
Boyer-Lindquist radial coordinate on $\lambda(s)$. Using Eq. (2), we
then show that this tetrad component of the Weyl tensor must
diverge along $\lambda(s)$ at least as fast as $\sim s^{-2}$, where $s$
is the affine parameter. This implies, by Proposition 2.1, that
$\lambda(s)$ must then terminate in Tipler's strong curvature
singularity. The rigorous proof is as follows.

\vspace{4mm}
{\em Proof}. Let us assume that the spacetime is parametrized by the
Boyer-Lindquist coordinates $(r,\theta,\phi,t)$. Since $\lambda(s)$
reaches the ring singularity at $r=0$, it must lie entirely in the
equatorial plane $\theta=\pi/2$ on the positive $r$ side. The
coordinate $r$ will change along $\lambda(s)$ according to Eq. (2).
{}From this equation it follows easily that there may exist at most two
positive values of $r$ on $\lambda(s)$ for which $dr/ds=0$, and so we
will have $dr/ds\neq 0$ along $\lambda(s)$ for all $r$ in a
sufficiently small interval about 0. It is thus clear that there must
exist some affine parameter value $s_1>0$ such that $dr/ds>0$ along
$\lambda(s)$ for all $s\in (0,s_1]$, because $r>0$ and $s>0$ on
$\lambda(s)$, and $r\rightarrow 0$ along $\lambda(s)$ as $s\rightarrow
0$. Let us now fix such a parameter value $s_{1}$. According to
Proposition 2.1, in order to prove the theorem, it suffices to show
that the rate of growth of the curvature along $\lambda(s)$ is strong
enough for $s$ in some small interval about 0. For convenience, and
without loss of generality, we can thus assume that the affine
parameter on $\lambda(s)$ ranges over the interval $(0,s_{1}]$.

Let ${\bf u}$ be the tangent vector to $\lambda(s)$. The coordinate
basis components $u^{r}$, $u^{\phi}$ and $u^{t}$ of ${\bf u}$ are given
by Eqs. (2)-(4), with the $+$ sign in Eq. (2) as $dr/ds>0$ along
$\lambda(s)$. The corresponding component $u^{\theta}$ of ${\bf u}$
identically vanishes as $\lambda(s)$ lies entirely in the equatorial
plane. Given any point $p\in \lambda(s)$, we define ${\bf k}$ to be the
vector in the tangent space $T_{p}$ with the following coordinate basis
components: $k^{r}=k^{\phi}=k^{t}=0$ and $k^{\theta}=r^{-1}$, where $r$
is the radial coordinate of $p$. One can readily verify that, with
respect to the scalar product given by the Kerr metric (1), ${\bf k}$
is a unit spacelike vector orthogonal to ${\bf u}$, i.e. ${\bf k\cdot
k}=1$ and ${\bf k\cdot u}=0$. In addition, ${\bf k}$ is parallely
transported along $\lambda(s)$ (see Appendix A). The pair $\{{\bf
k},{\bf u}\}$ can easily be extended to the pseudo-orthonormal tetrad
$\{{\bf E}_{i}\}$ mentioned in Proposition 2.1. To see this, let us
first fix some point $q\in \lambda(s)$. Let us now take a unit
spacelike vector orthogonal to the plane spanned by the vectors ${\bf
u}$ and ${\bf k}$ in the tangent space $T_{q}$; we will denote this
vector by ${\bf n}$. (Clearly, such a vector can always be found since
${\bf u}$ is null and ${\bf k}$ is spacelike.) Now let $\{{\bf
e}_{i}\}$ $(i=1,2,3,4)$ be some orthonormal basis in the space $T_{q}$,
with the spacelike vectors ${\bf e}_{2}$ and ${\bf e}_{3}$ chosen so
that ${\bf e}_{2}={\bf k}$ and ${\bf e}_{3}={\bf n}$. Using the
spacelike vector ${\bf e}_{1}$ and the timelike vector ${\bf e}_{4}$,
we now define ${\bf v}_{1}\equiv \alpha({\bf e}_{4}+{\bf e}_{1})$ and
${\bf v}_{2}\equiv \alpha({\bf e}_{4}-{\bf e}_{1})$, where $\alpha\neq
0$ is some constant. Evidently, these new vectors are null and each of
them is orthogonal to both ${\bf k}$ and ${\bf n}$. Moreover, as ${\bf
v}_{1}{\bf \cdot v}_{2}=-2\alpha^{2} \neq 0$, at least one of them, say
${\bf v}_{1}$, is not parallel to the vector ${\bf u}$, i.e. we must
have ${\bf u\cdot v}_{1}\neq 0$. By suitable choice of the constant
$\alpha$ in the definition of ${\bf v}_{1}$ one can always normalize
${\bf v}_{1}$ so that ${\bf u\cdot v}_{1}=-1$. The pseudo-orthonormal
tetrad $\{{\bf E}_{i}\}$ in the space $T_{q}$ can now be chosen as
follows: ${\bf E}_{1}\equiv {\bf n}$, ${\bf E}_{2}\equiv {\bf k}$,
${\bf E}_{3} \equiv {\bf v_{1}}$ and ${\bf E}_{4}\equiv {\bf u}$. By
parallely transporting this tetrad along $\lambda(s)$ one obtains the
pseudo-orthonormal basis at each point of $\lambda(s)$.

We shall now find one of the components $C^{i}\null_{jkl}$ of the
Weyl tensor with respect to the tetrad $\{{\bf E}_{i}\}$.
The components $C_{mjkl}$ of the Weyl tensor with respect to $\{{\bf
E}_{i}\}$ can be obtained from the coordinate components $C_{abcd}$ of
the Weyl tensor according to
\be
C_{mjkl}=C_{abcd}E^{a}_{m}E^{b}_{j}E^{c}_{k}E^{d}_{l},
\ee
where $E^{a}_{m}$ (resp., $E^{b}_{j}$, $E^{c}_{k}$ and $E^{d}_{l}$) is
the $a$th (resp., $b$th, $c$th and $d$th) coordinate basis component of
the vector ${\bf E}_{m}$ (resp., ${\bf E}_{j}$, ${\bf E}_{k}$ and ${\bf
E}_{l}$) of the tetrad $\{{\bf E}_{i}\}$. Having the tetrad components
$C_{mjkl}$, we can now easily find the tetrad components
$C^{i}\null_{jkl}$ of the Weyl tensor:
\be
C^{i}\null_{jkl}=\eta^{im}C_{mjkl},
\ee
where $\eta^{im}$ is the inverse of the matrix $\eta_{im}\equiv {\bf
E}_{i}\cdot {\bf E}_{m}$; that is, we have $\eta^{11}=\eta^{22}=
-\eta^{34}=-\eta^{43}=1$ and $\eta^{im}=0$ in all other cases. Since
the tetrad $\{{\bf E}_{i}\}$ is chosen so that ${\bf E}_{2}={\bf k}$
and ${\bf E}_{4}={\bf u}$, and the vectors ${\bf k}$ and ${\bf u}$ are
given in the explicit form, we can find, applying (5) and (6), an
explicit expression for the component $C^{2}\null_{424}$ of the Weyl
tensor with respect to $\{{\bf E}_{i}\}$. This is done in Appendix B;
the result is
\be
C^{2}\null_{424} = \frac{3m(L-aE)^2}{r^5}.
\ee

The task is now to show that if $\lambda(s)$ is not a member of the
principal null congruences, then there exists some affine parameter
interval $(0,s_0]$ of $\lambda(s)$ on which $C^{2}\null_{424}\geq
Ks^{-2}$, where $K$ is some positive constant. To do this, let us
first recall that the affine parameter $s$ and the radial coordinate
$r$ on $\lambda(s)$ are related by Eq. (2), with the + sign as
$dr/ds>0$ on $(0,s_1]$. Let us now rewrite this equation in the form
\be
\frac{ds}{dr} =  \frac{r^{3/2}}{F(r)},
\ee
where $F(r)\equiv \left[r^3E^2-r(L^2-a^2E^2)+2m(L-aE)^2\right]^{1/2}$.
As $dr/ds>0$ on $(0,s_1]$, it is obvious that $ds/dr>0$ on $(0,r_1]$,
where $r_1$ denotes the value of the coordinate $r$ of the point
$\lambda(s_1)$. By (8) it is thus clear that $F(r)$ is strictly
positive on $(0,r_1]$. It is also clear, as $\lim_{r\rightarrow
0}F(r)=|L-aE|\sqrt{2m}$, that $F(r)$ is bounded on $(0,r_1]$. So there
exists a positive number $F_0\equiv \sup\{F(r)|0<r\leq r_1\}$. Consider
now the function $y(r)\equiv 2r^{5/2}(5F_0)^{-1}$ defined on $(0,r_1]$.
Since $dy/dr=r^{3/2}/F_0$ and $F_0\geq F(r)>0$ on $(0,r_1]$, by (8) we
have $ds/dr|_{r'}\geq dy/dr|_{r'}>0$ for all $r'\in (0,r_1]$. From this
it follows easily, as $\lim_{r\rightarrow 0}y(r)=\lim_{r\rightarrow
0}s(r)=0$, that $s(r)\geq y(r)$ for all $r\in (0,r_1]$. Evidently, both
$s(r)$ and $y(r)$ are strictly increasing on $(0,r_1]$, and so there
exist their inverse functions. Let $r:(0,s_1]\rightarrow (0,r_1]$ be
the inverse function of $s(r)$ and let $z:(0,s_0]\rightarrow (0,r_1]$
be the inverse function of $y(r)$ (note that $s_0\leq s_1$ since
$s(r)\geq y(r)$ on $(0,r_1]$). From the fact that $s(r)\geq y(r)$ for
all $r\in (0,r_1]$, it follows immediately that $r(s)\leq z(s)$ for all
$s\in (0,s_0]$. Note also that $z(s)=(5F_0 s/2)^{2/5}$, which is clear
{}from the definition of $y(r)$. We thus have $r(s)\leq (5F_0 s/2)^{2/5}$
on $(0,s_0]$. Combining this inequality with (7), and taking into
account the fact that $m>0$, we obtain
\be
C^2\null_{424}=\frac{3m(L-aE)^2}{r^5(s)}\geq Ks^{-2}
\ee
for all $s\in (0,s_0]$, where $K\equiv 12m(L-aE)^2(5F_0)^{-2}$. Suppose
now that $\lambda(s)$ does not belong to either of the two principal
null congruences; then we have $L\neq aE$ (see Sec. 2), and hence
$K>0$. By (9) and Proposition 2.1 it is thus clear that $\lambda(s)$
must terminate in Tipler's strong curvature singularity as
$s\rightarrow 0$, which is the desired conclusion.

\section{Concluding remarks}

We have examined the curvature strength of the Kerr singularity with
$m>0$ and $a\neq~0$. We have shown that every null geodesic reaching
this singularity, with the exception of those belonging to the
principal null congruences, must in fact terminate in Tipler's strong
curvature singularity. This is a {\em typical} property of null
geodesics approaching the Kerr singularity because the principal null
geodesics are very special and can be considered to form a set of
``measure zero'' in the family of all null geodesics reaching the
singularity (Ref. \cite{chandra}, pp. 328, 329). The existence of these
special geodesics is due to the high symmetry of the solution and one
can expect that such geodesics will not occur in more general
spacetimes. Thus if one attempts to define the curvature strength of a
Kerr-like naked singularity, which could possibly arise in a {\em
generic} collapse, one way of doing this is to assume that {\em all}
null geodesics approaching such a singularity will behave much the same
as a typical null geodesic approaching the Kerr singularity with
$|a|>m$ ---i.e. will also terminate in a singularity of the strong
curvature type. Using this assumption, one may then attempt to
formulate and prove a theorem which would constraint or prohibit the
occurrence of Kerr-like naked singularities in generic collapse
situations. A theorem of this type was established in Ref.
\cite{rudnicki96}.

In this context, it is worth recalling that the inextendibility
condition assumed in the theorem of Ref. \cite{rudnicki96} may in fact
hold for a much more general class of possible singularities than only
those of the strong curvature type, for the curvature need not
necessarily diverge along geodesics satisfying this condition
\cite{rudzie}. However, we have checked that this condition fails
{}to hold for the principal null geodesics approaching the Kerr
singularity (in fact, it will always fail to hold for principal null
geodesics in any Ricci-flat spacetime). Thus the theorem of Ref.
\cite{rudnicki96} does not exclude the possibility that a naked Kerr
singularity accompanied by closed timelike curves could develop from
some non-singular initial data. However, from the proof of that theorem
it may be concluded that this can happen only if the inextendibility
condition fails to hold for {\em all} null geodesics terminating in the
past at the naked Kerr singularity and generating the future Cauchy
horizon due to the formation of this singularity. According to our
result, this is possible only if all these geodesics belong to the
principal null congruences. This would be a very special case.

\section*{Acknowledgments}
We wish to thank the referee for helpful comments.
This work was supported by the Polish Committee for
Scientific Research (KBN) under grant No. 2 P03B 073 15.

\appendix
\section{}

In this appendix, we demonstrate that the vector ${\bf k}$ is parallely
transported along the null geodesic $\lambda(s)$ with the tangent
vector ${\bf u}$. To this end we only need to show that
\begin{equation} \label{appa:1}
\left(\frac{\partial k^{c}}{\partial x^b} +
\Gamma^{c}\null_{ab}k^{a}\right)u^{b}=0,
\end{equation}
where $\Gamma^{c}\null_{ab}$ are the connection coefficients, which can
be obtained from the metric tensor $g_{ab}$ according to
\begin{equation} \label{appa:2}
\Gamma^{c}\null_{ab}=\frac{1}{2}g^{cd}\left(\frac{\partial
g_{bd}}{\partial x^a} +\frac{\partial g_{da}}{\partial x^b} -
\frac{\partial g_{ab}}{\partial x^d}\right).
\end{equation}
Since $u^\2=k^\1=k^\3=k^\4=0$ and $k^\2=r^{-1}$, the components of
Eq.~(\ref{appa:1}) in the $(\1,\2,\3,\4)$ coordinate system take the form \\
for $c=\1$:
\begin{equation} \label{appa:3}
\left(\Gamma^{\1}\null_{\2\1}u^\1 + \Gamma^{\1}\null_{\2\3}u^\3 +
\Gamma^{\1}\null_{\2\4}u^\4 \right)r^{-1} = 0,
\end{equation}
for $c=\2$:
\begin{equation} \label{appa:4}
-r^{-2}u^\1 + \left(\Gamma^{\2}\null_{\2\1}u^\1 +
\Gamma^{\2}\null_{\2\3}u^\3 +\Gamma^{\2}\null_{\2\4}u^\4 \right)r^{-1}
=0,
\end{equation}
for $c=\3$:
\begin{equation} \label{appa:5}
\left(\Gamma^{\3}\null_{\2\1}u^\1 +\Gamma^{\3}\null_{\2\3}u^\3 +
\Gamma^{\3}\null_{\2\4}u^\4 \right)r^{-1} = 0,
\end{equation}
for $c=\4$:
\begin{equation} \label{appa:6}
\left(\Gamma^{\4}\null_{\2\4}u^\1 +\Gamma^{\4}\null_{\2\3}u^\3 +
\Gamma^{\4}\null_{\2\4}u^\4 \right)r^{-1} = 0.
\end{equation}
Applying (\ref{appa:2}) to the Kerr metric (1), we can now calculate the
connection coefficients appearing in
Eqs. (\ref{appa:3})-(\ref{appa:6}); the result is
$$
\Gamma^{\1}\null_{\2\1} = -\frac{a^2 \sin\theta
\cos\theta}{\rho^2},\hspace{2em}
\Gamma^{\2}\null_{\2\1} = \frac{r}{\rho^2},\hspace{2em}
\Gamma^{\3}\null_{\2\3} =
      \frac{(\rho^2+2a^2mr\sin^2\theta)\cos\theta}{\rho^4\sin\theta},
$$
$$
\Gamma^{\3}\null_{\2\4} =
-\frac{2mar\cos\theta}{\rho^4\sin\theta},\hspace{2em}
\Gamma^{\4}\null_{\2\3} =
\frac{2ma^3r\sin^3\theta\cos\theta}{\rho^4},\hspace{2em}
\Gamma^{\4}\null_{\2\4} = \frac{2ma^2r\sin\theta\cos\theta}{\rho^4},
$$
$$
\Gamma^{\1}\null_{\2\3} = \Gamma^{\1}\null_{\2\4} =
\Gamma^{\2}\null_{\2\3} = \Gamma^{\2}\null_{\2\4} =
\Gamma^{\4}\null_{\2\1} = \Gamma^{\1}\null_{\2\3} = 0.
$$
Substituting these coefficients into Eqs.
(\ref{appa:3})-(\ref{appa:6}), and putting
$\theta=\pi/2$ as $\lambda(s)$ lies in the equatorial plane, we can now
readily see that Eqs. (\ref{appa:3})-(\ref{appa:6}) are satisfied,
as desired.

\section{}

We give in this appendix some details of the computation of the
component $C^{2}\null_{424}$ of the Weyl tensor with respect to
the tetrad $\{{\bf E}_{i}\}$. Since ${\bf E}_{2}={\bf k}$ and ${\bf
E}_{4}={\bf u}$, and $u^\2=k^\1=k^\3=k^\4=0$ and $k^\2=r^{-1}$, the
expression (5) for $C_{2424}$ takes the form
$$
C_{2424}=r^{-2}\left[C_{\2\1\2\1}(u^\1)^2+C_{\2\3\2\3}(u^\3)^2
+C_{\2\4\2\4}(u^\4)^2\right]
$$
\be \label{appb:1}
+2r^{-2}\left(C_{\2\1\2\3}u^\1u^\3+
C_{\2\1\2\4}u^\1u^\4+C_{\2\3\2\4}u^\3u^\4\right),
\ee
where the components $u^\1$, $u^\3$ and $u^\4$ of ${\bf u}$ are given
by Eqs. (2)-(4), with the + sign in Eq. (2). We recall that the Weyl
tensor $C_{abcd}$ is defined by
$$
C_{abcd} = R_{abcd} + g_{a[d}R_{c]b} + g_{b[c}R_{d]a}
	 + \frac{1}{3} R g_{a[c}g_{d]b},
$$
where $R_{abcd}$, $R_{ab}$ and $R$ denote the curvature tensor, the
Ricci tensor and the curvature scalar, respectively. Since the Kerr
spacetime is Ricci flat, $R_{ab}$ and $R$ will vanish, and so
$C_{abcd}=R_{abcd}$. In this case the coordinate components of the Weyl
tensor can be obtained from the metric components $g_{ab}$ according to
$$
C_{abcd} = \frac{1}{2}\left(\frac{\partial^2g_{ac}}{\partial x^d
\partial x^b}+ \frac{\partial^2g_{bd}}{\partial x^c \partial x^a} -
\frac{\partial^2g_{bc}}{\partial x^d \partial x^a} -
\frac{\partial^2g_{ad}}{\partial x^c \partial x^b}\right)
+g_{ef}(\Gamma^e\null_{ca}\Gamma^f\null_{bd}
-\Gamma^e\null_{da}\Gamma^f\null_{bc}),
$$
where the connection coefficients are given by (\ref{appa:2}). Applying this
formula to the Kerr metric (1), we can now calculate the coordinate
components of the Weyl tensor appearing in (\ref{appb:1}); the result is
$$
C_{\2\1\2\1 } =
  \frac {mr (3a^2\cos^2\theta -r^2)}{\rho^2\Delta},
$$
$$
C_{\2\3\2\3} = -\frac {mr (3a^2 \cos^2 \theta - r^2 )
[a^2\Delta \sin^2 \theta +2(r^2+a^2)^2] \sin^2 \theta}{\rho^6},
$$
$$
C_{\2\4\2\4} = -\frac{mr (3a^2 \cos^2 \theta -r^2)(2a^2
\sin^2\theta+\Delta)}{\rho^6},
$$
$$
C_{\2\3\2\4} =\frac {mra (3a^2 \cos^2\theta-r^2) [\Delta+2(r^2+a^2)]
\sin^2\theta }{\rho^6},
$$
$$
C_{\2\1\2\3} = C_{\2\1\2\4} = 0.
$$
Inserting these expressions in (\ref{appb:1}), and putting
$\theta=\pi/2$ as $\lambda(s)$ lies in the equatorial plane, we get
$$
C_{2424} = \frac{3m(L-aE)^2}{r^5}.
$$
Finally, we note that $C_{2424}=C^2\null_{424}$, which is clear from
(6).

\end{document}